\documentclass[aps,pre,twocolumn,groupedaddress]{revtex4}

\usepackage[dvips]{graphicx}

\begin{document}


\title{Diffuse-interface model  
for rapid phase transformations in nonequilibrium systems}

\author{Peter Galenko}
\email[]{e-mail:  Peter.Galenko@dlr.de;  Fax: ++49-(2203)-6012255}
\affiliation{Institut f\"ur Raumsimulation, DLR, K\"oln, D-51170, Deutschland}
\author{David Jou}
\affiliation{Departament de F\'isica, 
Universitat Aut\`onoma de Barcelona, 08193 Bellaterra, Catalonia, Spain}
\date{\today}

\begin{abstract}
A thermodynamic approach to rapid phase transformations 
within a diffuse interface in a binary system is developed. 
Assuming an extended set of independent thermodynamic 
variables formed by the union of the classic set of slow 
variables and the space of fast variables, we introduce  
finiteness of the heat and solute diffusive propagation at 
the finite speed of the interface advancing. 
To describe the transformation within the diffuse interface, 
we use the phase-field model which allows us 
to follow the steep but smooth change of phases  
within the width of diffuse interface. 
The governing equations of the phase-field model are derived for 
the hyperbolic model, model with memory, 
and for a model of nonlinear evolution of transformation within the diffuse-interface. 
The consistency of the model is proved by the condition 
of positive entropy production and by the outcomes of the  
fluctuation-dissipation theorem. A comparison with the existing 
sharp-interface and diffuse-interface versions of the model is given. \\
\\
\begin{pacs}
\pacs{PACS numbers: 05.70.Fh; 05.70.Ln; 64.60.-i; 83.20.Hn}
\end{pacs}

\end{abstract}



\keywords{Irreversible thermodynamics, phase transformation, 
	diffuse interface, phase field}

\maketitle
\section{Introduction}\label{sec:intro}
Arising from phase transformations, a classic free-boundary problem  
introduces a model of phase interface with zero thickness. 
Within this problem, a sharp discontinuity in properties or a jump 
of fluxes and thermodynamic functions occurs across the interface. 
The sharp-interface model has been a successful description 
of many physical phenomena in various systems \cite{friedman}. 
However, the sharp-interface model has difficulties 
in describing situations when interfacial thickness 
becomes comparable with the characteristic length of the considered 
phenomenon, and when a topology of the interface becomes 
complicated or multiply connected. In order to avoid these 
difficulties in the sharp-interface model, an alternative model 
with a finite interfacial thickness was suggested for explaining 
phase transformations \cite{caginalp}. 

Historically, the first formulation of basic principles 
of diffuse interfaces was given by Poisson, Maxwell, 
and Gibbs \cite{poisson} who suggested to consider  
interface as a region with finite thickness in which 
a steep but smooth transition of physical properties 
of phases occurs. Lord Rayleigh, van der Waals, and 
Korteweg \cite{rayleigh} applied thermodynamical 
principles to develop gradient theories for the 
interfaces with non-zero thickness. Through the 
past century, ideas of diffuse interface given by 
these authors \cite{poisson,rayleigh} were refined 
and applied in many physical phenomena 
(see, e.g., overviews in Ref.~\cite{row}). 

The diffuse-interface formalism has been widely 
applied to phase transformations in condensed 
media. Borrowing 
the formalism of the Landau theory of phase transitions 
\cite{landau_00}, the first introduction of the diffuse 
interface into the theory of phase transformations 
was made by Landau and Khalatnikov \cite{landau} 
who labelled the different phases by an additional 
order parameter to describe anomalous sound 
absorption of liquid helium. In its well-known form, 
the formal variational approach 
was established by Ginzburg and Landau for 
phase transitions from the normal to the 
superconducting phase \cite{ginz}. 
On the basis of this approach, 
diffuse-interface models with order parameters 
have been developed by Halperin, 
Hohenberg and Ma for the theory of critical 
phenomena \cite{C-model}, and by Allen and Cahn for 
antiphase domain coarsening \cite{ch3}.  

The diffuse-interface model has been also developed for the 
description of phase transformations of the first order, especially, 
for the solidification phenomenon. The diffuse-interface model of solidification 
incorporates an order parameter in the form of a phase-field variable \cite{fix}. 
The phase-field $\Phi$ has a constant value in homogeneous phases, e.g. $\Phi=-1$ 
for unstable liquid phase which is transforming into the solid phase 
with $\Phi=+1$. Between these phases in the interfacial region, 
the phase field, $\Phi$, changes steeply but smoothly from 
$-1$ to $+1$. In numerical solutions, it allows one to avoid explicit 
tracking of the interface and locate the interface at $\Phi=0$ \cite{sahm}. 
As a particular case, 
the phase-field model is reduced to the sharp interface limits 
\cite{caginalp1} and adopts the major models of sharp interface 
(such as Hele-Shaw type models, classical or modified 
Stefan problem, etc.). The phase-field 
$\Phi$ is considered as an order parameter which is 
introduced to describe the moving interfacial boundary 
between the initially unstable phase and the final phase.

Several thermodynamically consistent phase-field models 
have been proposed \cite{fife1,bis,anderson,garke}.
These include models for transformation in a pure system \cite{fife1} 
up to the rather general modelling of multiphase transformation 
in a multi-component system \cite{garke}. All of these models assume local equilibrium 
in the system, being consistent with the basic hypothesis 
of the classic irreversible thermodynamics (CIT) \cite{op,grmaz}. 
This assumption leads to the examination of a number of transport  
processes with small and moderate deviations from thermodynamic 
equilibrium and, as a consequence, relatively slow movement of the 
interface can be predicted. In principle, such an approach can be 
extended to the case when the condition of equilibrium is violated locally 
at the interface, e.g., as it has been made for solute trapping 
and kinetic effects \cite{wheel201}. However, the local equilibrium 
is missing both at the interface and within the bulk phases 
for rapid transformations such as rapid solidification \cite{gs}. 
In this case, the description of rapid phase trasformations might 
be provided by the formalism of extended irreversible thermodynamics 
(EIT) \cite{j1} 
which gives a causal description of transport processes and abandons the 
assumption of local equilibrium. An extension 
of the phase-field methodology for rapid transformation, which is 
caused by significant deviations from thermodynamic equilibrium, 
has been made recently \cite{g1}. 

The main purpose of the present paper is to describe a thermodynamically 
consistent model for rapid phase transformation in a binary system under 
local nonequilibrium conditions. Using the phase-field methodology, 
we derive the governing equations compatible with the macroscopic 
formalism of EIT and the microscopic fluctuation-dissipation theorem. 

The paper is organized as follows. In Sec. \ref{sec:mod}, 
a thermodynamic description of a considered system is given. 
We introduce the dissipative 
diffusion fluxes for heat and mass transport together with the 
phase-field rate of change as independent variables. 
In Sec. \ref{sec:ea}, the generalized Gibbs equation and entropy 
balance applicable to rapid advancing of the diffuse interfaces 
are given. 
As a starting point of the present phase-field model, an entropy 
functional is used in Sec. \ref{sec:hyper_pfm}. 
The analysis of the present phase-field model leads to the 
governing equations for the hyperbolic system with dissipation. 
In Sec. \ref{sec:generalizat}, a generalization of the hyperbolic phase-field 
model is given using the flux relaxation functions as well as 
a variational principle. In Sec. \ref{sec:comparison}, 
the model equations are compared with the 
outcomes of the existing sharp-interface and diffuse-interface models. 
Finally, in Sec. \ref{sec:con} we present a summary of our conclusions. 

\section{Description of the system}\label{sec:mod} 
\subsection{Thermodynamic variables}
Let us consider an isobaric binary system at 
nonuniform temperature $T$ with no convective flow 
and given concentration of atoms $A$ and $B$. 
The local equilibrium hypothesis establishes that the local and 
instantaneous correlations among the properties of the system 
are the same as for the whole system 
at a global equilibrium. Describing the nonequilibrium system 
as an ensemble of small local volumes in internal equilibrium, 
CIT~\cite{grmaz} is applicable to processes 
not too far from equilibrium. In addition to CIT, a local 
nonequilibrium formalism applicable to the 
strongly nonequilibrium systems has been 
developed in past two decades 
\cite{j1,jp1,j,mr}. As a phenomenological 
theory, this formalism is well-known as EIT \cite{j1,j2} which   
goes beyond the hypothesis of  local equilibrium and avoids 
the paradox of propagation of disturbances with an infinite 
speed. 

A fundamental problem in attempting to describe systems out of equilibrium 
is to select the relevant variables needed for a valid description 
of a nonequilibrium state. This problem has been 
intensively discussed in the literature (see references in 
bibliographic overview \cite{jcvl1}). A selection of the basic 
state space with the inclusion of the dissipative fluxes 
is formulated in EIT \cite{j} and tested against 
experimental data \cite{luz}. 
Accordingly, we extend the classic set of independent 
thermodynamic variables by the inclusion of dissipative fluxes as 
additional basic variables. 

CIT is based on the local equilibrium hypothesis \cite{op,grmaz} 
which assumes instant relaxation of fluxes to their steady-state 
values and describes the ensemble of atoms within local 
volumes by the Gibbs-Boltzmann statistics. In the standard formalism of the 
diffuse-interface using CIT, the set $\{C\}$ of independent  
variables is assumed to consist of the conserved variables such as 
energy density $e(\vec r,t)$ and concentration $X(\vec r,t)=X_B/(X_A+X_B)$ 
of the $B$ component in the system, and the 
non-conserved phase-field $\Phi(\vec r,t)$ variable 
[where $t$ is the time, and $\vec r$ is the position-vector of a point within system]. 
This can be expressed formally as follows: $\{C\} = \{e, X, \Phi \}$. 

The extended space of independent variables $\mathbf{E}$ is formed by 
the union of the classical set $\{C\}$ and the additional space $\{F\}$ 
of the fluxes of heat $\vec q$ and solute $\vec J$, and also 
the rate of change $\partial \Phi / \partial t$ of the phase-field variable, i.e. 
$\{F\} = \{\vec q, \vec J, \partial \Phi / \partial t \}$.  
This yields 
\begin{eqnarray}
\mathbf{E} = \{C\}~\cup~\{F \} 
= \{e,X,\Phi\}~\cup~\{\vec q, \vec J, \partial \Phi / \partial t \}. 
\label{A3}
\end{eqnarray}
Here $\{F\}$ is the space of fast non-conserved 
thermodynamic variables. 

There are, in fact, different possible choices of variables 
(fluxes in EIT, microstructural details in theories with internal variables), 
and the specific choice to be adopted depends on the aims of the 
description and on the problems to be analyzed. 
This does not mean that different choices of variables 
are incompatible with each other. For instance, 
in the study of flowing polymer solutions one may 
select as independent variables either the viscous 
pressure tensor or the conformation tensor describing 
the average microstructure of the macromolecules of 
the system; a Legendre transform exists which allows 
one to pass from one description to the other, 
in a similar way as it is possible, in equilibrium thermodynamics, 
to pass from a description using internal energy 
as independent variable to a description using 
absolute temperature as an independent variable \cite{jou2000}. 

Thus, our choice of the fluxes as variables does not exclude 
other possibilities. To justify our choice, we  
comment on the qualitative grounds, the meaning and the relevance of 
$\vec q$, $\vec J$, and $\partial \Phi /\partial t$ as variables. 
The fluxes $\vec q$ and $\vec J$ describe the 
exchanges of heat and matter between the interface and the 
neighbouring bulk phases. The fluxes do not instantaneously 
follow by the classical Fourier and Fick laws, relating them 
with temperature and concentration gradients, it takes them some 
time (usually rather short) to reach the value predicted by the 
classical transport equations. Obviously, when the interface 
motion is fast enough, the delay effects in the dynamics 
of the fluxes may play a determining role. This happens, 
for instance, when the velocity $V$ of the interface becomes 
comparable or higher than $l/\tau$, $l$ being the mean-free-path 
of the particles and $\tau$ the relaxation time of the fluxes. 
Thus, in these circumstances, $\vec q$ and $\vec J$ behave 
as independent variables with their own dynamics, which has 
important consequences on the dynamics and stability 
of the interface \cite{gd3,gd4}. 

The introduction of $\partial \Phi/\partial t$ as a further independent variable 
is motivated by similar, though slightly different consideration. Indeed, the 
space variation of $\Phi$ is related, among other factors, 
to the width of the interface. Thus, including $\partial \Phi/\partial t$ 
as an independent variable allows for a more detailed description 
of the internal kinetics and shape of the interface. 
In the same way as in Newtonian mechanics, where the initial position 
and velocity of a particle must be specified to determine their 
evolution, here we take both $\Phi$ and $\partial \Phi/\partial t$ 
as independent variables. If inertial effects are sufficiently low in 
comparison with dissipative effects, $\partial \Phi /\partial t$ 
will be directly determined in terms of $\Phi$ and its gradient 
by a dynamical equation. Otherwise, 
$\Phi$ and $\partial \Phi /\partial t$ will be independent and an 
equation for $\partial^2 \Phi /\partial t^2$ must be found.

\begin{table*}
\caption{\label{tab:1}
Relaxation time for the fluxes of heat, solute diffusion, and phase-field}
\begin{ruledtabular}
\begin{tabular}{lllll}
    System & $ \tau_T$ (s) & $ \tau_D$ (s) & $ \tau_\Phi$ (s)  \\
\hline
    Carbon tetrachloride & $2.50 \cdot 10^{-13}  $, Ref.~\cite{net_60} & --- & ---  \\
    Benzene & $1.22\cdot 10^{-13}$, Ref.~\cite{net_60} & --- & --- &  \\
    Nickel & $1.20\cdot 10^{-11}$, Ref.~\cite{eval1} & --- & $2.30\cdot 10^{-11}$, Ref.~\cite{eval2}  \\
    Diluted alloy Ni - 0.7 at.\% B & --- & $1.54\cdot 10^{-11}$, Ref.~\cite{eval21} & --- \\
    Concentrated alloy Cu - 30 at.\% Ni  & --- & $0.75\cdot 10^{-11}$, Ref.~\cite{eval211}& $7.92\cdot 10^{-11}$, Ref.~\cite{eval3}  \\
\end{tabular}
\end{ruledtabular}\\
\end{table*}

Consequently, with taking the above choice of variables, 
one may distinguish between the two sets of independent 
variables as it follows. 
The variables from the set $\{C\}$ are 
characterized as the slow 
variables, as their behavior is governed by conservation laws 
for energy and solute concentration plus the evolution of 
the phase-field, and as they decay slowly in time. 
In contrast, 
the independent space $\{F\}$ consists of 
non-conserved variables with relatively high rate of decay. 
The variables from $\{F\}$ differ from their classical value during intervals 
of the order of magnitude of the characteristic times $\tau_i$ for relaxation 
of the heat flux, solute diffusion 
flux, and rate of change of the phase-field variable. 
For time intervals much longer than these relaxation times $\tau_i$, the rate of 
variation of the fluxes can be ignored. 

\subsection{Relaxation times} \label{sec:rt} 
Generally, the relaxation times $\tau_i$ represent physically reasonable 
time estimations for the spontaneous return of the system 
to the steady state after some sudden perturbation. 
The relaxation times $\tau_T$ and $\tau_D$ for the heat and solute 
can be considered as the times needed for smoothing of 
inhomogeneities of temperature and concentration, 
respectively, by diffusion. The time $\tau_\Phi$ of relaxation 
for the phase-field can be evaluated from the velocity of the diffuse 
interface moving through the local volume with the characteristic spatial length. 
Consequently, the rate of decay of the heat flux $\vec q$, 
solute diffusive flux $\vec J$, and phase-field rate of change $\partial \Phi / \partial t$
are estimated by the following characteristic times 
\begin{eqnarray}
\tau_T=a/V_T^2, \qquad
 \tau_D=D/V_D^2, \qquad
   \tau_\Phi=l/V, \label{A1}
\end{eqnarray}
where $a$ is the thermal diffusivity, $V_T$  
the finite speed for heat diffusion (i.e. the speed of propagation of temperature 
disturbances), $D$ the solute diffusion constant, $V_D$  
the finite speed for diffusion (i.e. the speed of propagation of concentration 
disturbances), $V$ the velocity of the diffuse interface, and $l$  
the spatial length. 

For instance, the time $\tau_T$ is defined by the 
phonon-electron and phonon-phonon interactions 
for heat diffusion 
in metallic systems and it is estimated in Ref.~\cite{peierls1} 
to be in the range of  
$10^{-13} s <\tau_T<10^{-11} s$. 
The time $\tau_D$ is defined by the time for diffusion jumps 
of particles, which varies within 
a wide interval: $10^{-11} s <\tau_D<10^{-7} s$ in a binary 
alloy system or inorganic solution \cite{galenko1}. 
In addition to this, the time $\tau_\Phi$ might be evaluated 
numerically from Eq.~(\ref{A1}) assuming that the length 
$l=W_0$ the width of the diffuse interface and the 
velocity $V$ is the characteristic velocity for rapid adiabatic transformations. 
Thus, for numeric evaluation of $\tau_\Phi$ in a pure system, 
one may accept the following expression
\begin{eqnarray}
\tau_\Phi=W_0 \chi/(\mu_0 Q), \label{A_tau} 
\end{eqnarray}
where $Q$ is the heat of transformation, 
$\chi$ is the heat capacity (so that relation $Q/\chi$ is considered as 
the characteristic temperature for adiabatic transformation), and 
$\mu_0$ the coefficient for atomic kinetics. Taking the values for pure 
nickel, e.g., $Q/\chi=418$ K \cite{barth}, $\mu_0=0.52$ m/(s$\cdot$K) 
\cite{hoyt}, and $W_0=5\cdot 10^{-9}$ m, one gets 
$\tau_\Phi=2.30\cdot 10^{-11}$ s. This value for $\tau_\Phi$ fits well 
to the time of diffuse-interface kinetics which might be calculated from 
the ``thin-interface'' analyses of Karma and Rappel \cite{karma1} 
extended by Bragard et al. \cite{bragard}.  

It is also reasonable to evaluate the relaxation time for the 
phase-field in a binary system using outcomes of the phase-field 
model via ``thin-interface'' analyses of Karma and Rappel \cite{karma1}. 
Namely, for the nonisothermal solidification of a binary system, 
Ramirez et al. \cite{ramirez1} derived the time $\tau_\Phi$ for 
the phase-field as a function of $X$ and $\Phi$. It is described by 
\begin{eqnarray}
\tau_\Phi=\frac {W_0^2}{\Gamma} \left( \frac{1}{\mu_0} 
+ a_1 a_2 \frac {W_0}{D}
\left[ \frac{DQ}{a\chi}
+ \frac{m(1-k)X}{1+k-(1-k)\Phi}	\right]	\right). 
\label{TAU2}
\end{eqnarray}
For numeric evaluation, 
we accept  the following material parameters 
for a Cu-Ni metallic system in Eq. (\ref{TAU2}): 
diffuse-interface width $W_0=1\cdot 10^{-9}$ m, 
Gibbs-Thomson coefficient $\Gamma=1.3 \cdot 10^{-7}$ K$\cdot$m \cite{gd1}, 
atomic kinetics coefficient $\mu_0=0.24$ m/(s$\cdot$K) \cite{gd1}, 
constants $a_1=0.8839...$ and $a_2=0.6267...$ \cite{karma1}, 
solute diffusion constant $D=3 \cdot 10^{-9}$ $m^2$/s \cite{gd1}, 
thermal diffusivity $a=1.5 \cdot 10^{-5}$ $m^2$/s \cite{gd1},
adiabatic temperature (relation of latent heat and heat capacity) 
$Q/\chi=402$ K \cite{herl1}, 
slope of the liquidus line $m=4.38$ K/at.$\%$ \cite{herl1}, 
solute partitioning coefficient $k=0.81$ \cite{herl1}. 
As a result following from Eq. (\ref{TAU2}), 
one gets $\tau_\Phi=7.92\cdot 10^{-11}$ s for the values of 
$X=70$ at.$\%$ and $\Phi=0.5$.

The values for the relaxation 
times for some pure and binary systems are summarized 
in Table \ref{tab:1}. It can be seen, e.g. for metals and alloys, 
that even though 
the heat speed $V_T$ is much larger than the solute diffusion speed 
$V_D$, the relaxation times for $\vec q$ and $\vec J$ have the same 
order of magnitude, i.e. $\tau_T \approx \tau_D$. Therefore, a front of 
the heat profile moves with a speed much higher than a front 
of the solute diffusive profile. However, due to the fast thermal diffusion, 
$a>>D$, the relaxation of the heat flux $\vec q$ proceeds  
approximately at the same characteristic time as the relaxation for 
solute diffusion flux $\vec J$.

\section{Entropy approach}\label{sec:ea} 
\subsection{Generalized Gibbs equation}
For the local nonequilibrium system described in Sec. \ref{sec:mod}, 
we postulate the existence of a local generalized 
entropy density $s$ whose set of variables is the 
extended space $\mathbf{E}$ by Eq.~(\ref{A3}). 
The generalized Gibbs equation for $s$ is described by 
\begin{eqnarray}
& & ds(e, X, \Phi, \vec q, \vec J, \partial \Phi / \partial t ) = 
ds_e(e, X, \Phi) \nonumber\\ 
\nonumber\\ 
& & + ds_{ne} (\vec q, \vec J, \partial \Phi/ \partial t) 
= \frac{\partial s_e}{\partial e}de + \frac{\partial s_e}{\partial X}dX 
+ \frac{\partial s_e}{\partial \Phi}d\Phi \nonumber\\ 
\nonumber\\ 
& & + \frac{\partial s_{ne}}{\partial \vec q}\cdot d\vec q + \frac{\partial s_{ne}}{\partial \vec J} \cdot d\vec J 
+ \frac{\partial s_{ne}}{\partial (\partial \Phi/\partial t)}d \left( \frac {\partial \Phi}{\partial t} \right).  
\label{B21}
\end{eqnarray}
In Eq.~(\ref{B21}), $s_e$ is a local equilibrium 
contribution defined on the set $\{C\}$ of the 
classic slow variables $\{e, X, \Phi \}$, and  
$s_{ne}$ is a flux-dependent purely nonequilibrium part of 
the generalized entropy defined on the space $\{F\}$ consisting 
of the fluxes $\{ \vec q, \vec J, \partial \Phi / \partial t \}$ 
as the independent fast variables. 

The derivatives of the entropy density with respect to 
classical variables and their fluxes appearing in Eq.~(\ref{B21}) 
are described by  
\begin{eqnarray}
& & \frac{\partial s_e}{\partial e} = \frac{1}{T},  \qquad
\frac{\partial s_e}{\partial X}= - \frac{\Delta \mu}{T},  \nonumber\\ 
\nonumber\\ 
& & \frac{\partial s_e}{\partial \Phi} 
= (1-X) \frac{\partial s_A}{\partial \Phi} 
+ X \frac{\partial s_B}{\partial \Phi}, \nonumber\\ 
\nonumber\\ 
& & \frac{\partial s_{ne}}{\partial \vec q} = - \alpha_q \vec q, \qquad
\frac{\partial s_{ne}}{\partial \vec J} = - \alpha_j \vec J, \nonumber\\ 
\nonumber\\ 
& & \frac{\partial s_{ne}}{\partial (\partial \Phi/\partial t)} 
= - \alpha_\phi \frac{\partial \Phi}{\partial t},  
\label{B5}
\end{eqnarray}
where $\Delta \mu = \mu_A - \mu_B$ is the difference of the chemical potentials 
for components $A$  and $B$, respectively, 
and $s_A$ and $s_B$ are the entropies for pure components $A$ and $B$, 
respectively. The chemical potentials and entropies of 
components can be chosen for every concrete system 
(see, e.g., Refs.~\cite{bis,garke}). 

In Eqs.~(\ref{B5}), the coefficients $\alpha_i$ are scalars which do not 
depend on $\vec q$, $\vec J$, and 
$\partial \Phi / \partial t$ and are assumed to be  
\begin{eqnarray}
& & \alpha_q = \left( \frac {\tau_T}{\kappa T^2} \right)_{X,\Phi},  \qquad 
\alpha_j = \frac {\tau_D}{TD} \left (\frac {\partial (\Delta \mu)}{\partial X} \right)_{T,\Phi}, \nonumber\\ 
\nonumber\\ 
& & \alpha_\phi = \left ( a_0\frac {\tau_\Phi W_0 Q}{T \mu_0} \right)_{T,X},
\label{1111}
\end{eqnarray}
where $\kappa$ is the thermal conductivity, 
$a_0$ a dimensionless factor (dependent on the model of the 
diffuse interface, specifically leading to the sharp-interface asymptotic 
limit), and $Q$ the heat of the transformation.

After integration, Eq.~(\ref{B21}) can be written in the form  
\begin{eqnarray}
& & s(e, X, \Phi, \vec q, \vec J, \partial \Phi / \partial t ) = s_e(e, X, \Phi) 
+ s_{ne} (\vec q, \vec J, \partial \Phi/ \partial t), \nonumber\\ 
\nonumber\\ 
& & s_{ne}(\vec q, \vec J, \partial \Phi / \partial t) = - \frac{\alpha_q}{2} \vec q \cdot \vec q 
- \frac {\alpha_j}{2} \vec J \cdot \vec J 
- \frac{\alpha_\phi}{2} \left( \frac {\partial \Phi}{\partial t} \right)^2. \nonumber\\
\label{0}
\end{eqnarray}
Consequently, we arrive to a generalized entropy density 
given by an expansion around its local equilibrium value 
up to second-order in the fluxes. 
In the limit of infinite speeds 
($V_T \rightarrow \infty$, $V_D \rightarrow \infty$, and $V \rightarrow \infty$), 
one gets $\tau_T \rightarrow 0$, $\tau_D \rightarrow 0$, and 
$\tau_\Phi \rightarrow 0$. In such a case, the term $s_{ne}$ 
vanishes and Eq.~(\ref{0}) gives the entropy 
density $s_e(e,X,\Phi)$ for local equilibrium system. 

\subsection{Entropy balance}
For the system described by the 
extended set $\mathbf{E}$ of variables, Eq.~(\ref{A3}), 
the local balance laws for the energy and concentration 
are given by
\begin{eqnarray} 
\frac{\partial e}{\partial t} + \nabla \cdot \vec q = 0, 
\qquad \frac{\partial X}{\partial t} + \nabla \cdot \vec J = 0, 
\label{B511}
\end{eqnarray}
and the evolution of entropy density is defined by 
\begin{eqnarray}
\frac{\partial s}{\partial t} + \nabla \cdot \vec J_S = \sigma_S.  
\label{15}
\end{eqnarray}
The change of the total entropy $S$ in time $t$ is described by 
\begin{eqnarray}
\frac {dS}{dt} = 
\left (\frac{dS}{dt} \right)_{ex} + \left (\frac{dS}{dt} \right)_{in}, 
\label{11} \end{eqnarray} 
where 
\begin{eqnarray}
\left (\frac{dS}{dt} \right)_{ex} = - 
\int_{v} \nabla \cdot \vec J_S dv = - 
\int_{\omega} \vec J_S \cdot \vec n d\omega, 
\label{12} 
\end{eqnarray}
is the external exchange of entropy due to entropy flux $\vec J_S$ and 
\begin{eqnarray}
\left (\frac{dS}{dt} \right)_{in} = 
\int_{v} \sigma_S dv,
\label{13}
\end{eqnarray}
is the internal production of entropy due to dissipation within the system. 
In Eqs.~(\ref{12}) and (\ref{13}): $\omega$ is the outer surface of sub-volume 
$v$, $\vec n$ the normal vector to the surface, 
and $ \sigma_S$ the local entropy production. 

\section{Hyperbolic phase-field model}\label{sec:hyper_pfm} 
In this section, the important class of hyperbolic models 
with dissipation is considered. We work out the explicit 
evolution equations for the variables including the relaxation terms.

\subsection{An entropy functional}\label{sec:entropy}
Now we use an entropy functional of the following form 
\begin{eqnarray}
& & S = \int_{v} \big[ s(e, X, \Phi, \vec q, \vec J, \partial \Phi / \partial t) \nonumber\\ 
\nonumber\\ 
& & - \frac {\varepsilon_e^2}{2}|\nabla e|^2 
- \frac {\varepsilon_x^2}{2}|\nabla X|^2 
- \frac {\varepsilon_\phi^2}{2}|\nabla \Phi|^2 \big] dv.
\label{1}
\end{eqnarray}
Here $\varepsilon_e$, $\varepsilon_x$, and $\varepsilon_\phi$ 
are constants for the energy, concentration, and phase-field, 
respectively. In the functional (\ref{1}) the gradient 
terms $|\nabla e|^2$, $|\nabla X|^2$, and $|\nabla \Phi|^2$ 
are used to describe spatial inhomogeneity within 
the fields according to previous diffuse-interface models 
\cite{ginz,ch3,sahm}. It is logical to include 
gradient terms in Eq. (\ref{1}) [of the so-called ``Ginzburg-Landau 
form''] because, as stressed before, our interest is focused on interfaces 
with steep gradients. In addition, the extension (\ref{A3}) 
gives the entropy density $s$ based also on the fluxes 
$\vec q$, $\vec J$, and $\partial \Phi /\partial t$ 
as independent variables. 

To obtain the evolution of the entropy, Eq.~(\ref{11}), and consider 
the several parts of the entropy exchange, Eqs.~(\ref{12})-(\ref{13}), 
we differentiate Eq.~(\ref{1}) with respect to time. Combining the terms,  
after some algebra one obtains 
\begin{widetext}
\begin{eqnarray}
& & \frac {dS}{dt} = \int_{v} \left[ \frac{\partial s}{\partial e} + 
\varepsilon_e^2 \nabla^2 e \right] \left(\frac {\partial e}{\partial t}\right) dv 
+ \int_{v} \left[ \frac{\partial s}{\partial X} + 
\varepsilon_x^2 \nabla^2 X \right] \left(\frac{\partial X}{\partial t}\right) dv 
+ \int_{v} \left[ \frac{\partial s}{\partial \Phi}  
+ \varepsilon_\phi^2 \nabla^2 \Phi \right] 
\left(\frac{\partial \Phi}{\partial t}\right) dv \nonumber\\ 
\nonumber\\ 
\nonumber\\ 
& & + \int_{v}  
 \left[ \frac{\partial s}{\partial \vec q} 
\left( \frac{\partial \vec q}{\partial t} \right) 
+ \frac{\partial s}{\partial \vec J} 
\left( \frac{\partial \vec J}{\partial t} \right) 
+ \frac{\partial s}{\partial (\partial \Phi / \partial t)}
\left( \frac{\partial^2 \Phi}{\partial t^2} \right) 
 \right]dv \nonumber\\ 
\nonumber\\ 
\nonumber\\ & & - \int_{\omega} 
\left[ \varepsilon_e^2 
\left(\frac{\partial e}{\partial t}\right) \nabla_n e + 
\varepsilon_x^2 \left(\frac{\partial X}{\partial t}\right) \nabla_n X 
+ \varepsilon_\phi^2 \left(\frac{\partial \Phi}{\partial t}\right) \nabla_n \Phi \right] d\omega, 
\label{3}
\end{eqnarray} 
\end{widetext}
where $\nabla_n$ is the gradient vector pointed by the normal vector 
$\vec n$. 

Now we substitute the balance laws for energy and concentration, 
Eqs.~(\ref{B511}), into Eq.~(\ref{3}), and then use the 
theorem of divergence. One gets 
\begin{widetext}
\begin{eqnarray} 
& & \frac{dS}{dt} = 
- \int_{\omega} \bigg\{ \varepsilon_e^2 
\left (\frac{\partial e}{\partial t}\right) \nabla_n e 
+ \left (\frac {\partial s}{\partial e}
+ \varepsilon_e^2 \nabla^2 e \right) q_n + \varepsilon_x^2 
\left(\frac{\partial X}{\partial t}\right) \nabla_n X + 
\left (\frac {\partial s}{\partial X}
+ \varepsilon_x^2 \nabla^2X \right) J_n + 
\varepsilon_\phi^2 \left(\frac{\partial \Phi}{\partial t}\right) \nabla_n \Phi 
\bigg\}d\omega \nonumber\\
\nonumber\\
\nonumber\\
& & + \int_{v} \bigg\{ \vec q \cdot \nabla \left [\frac {\partial s}{\partial e} 
+ \varepsilon_e^2 \nabla^2 e \right] + \frac {\partial s}{\partial \vec q} 
\frac {\partial \vec q}{\partial t} + 
\vec J \cdot \nabla \left [\frac {\partial s}{\partial X} 
+ \varepsilon_x^2 \nabla^2 X \right] + \frac {\partial s}{\partial \vec J} 
\frac {\partial \vec J}{\partial t}  
+ \frac{\partial \Phi}{\partial t}
\left[ \frac{\partial s}{\partial \Phi} +
\varepsilon_\phi^2 \nabla^2 \Phi \right] 
+ \frac{\partial s}{\partial (\partial \Phi / \partial t)}
\frac{\partial^2 \Phi}{\partial t^2}\bigg\}dv,  \nonumber\\ 
\label{6} \end{eqnarray} 
\end{widetext}
where $q_n$ and $J_n$ are the diffusion fluxes 
pointed by the normal vector $\vec n$. 

Using Eq.~(\ref{B5}), the change of the entropy, Eqs.~(\ref{11})-(\ref{13}), 
is obtained from Eq.~(\ref{6}). This yields 
\begin{eqnarray}
\frac{dS}{dt} = - \int_{\omega} J_S d\omega + \int_{v} \sigma_S dv, 
\label{8}
\end{eqnarray} 
where 
\begin{eqnarray}
& & J_S = \varepsilon_e^2 
\left(\frac{\partial e}{\partial t}\right) \nabla_n e + 
\left (\frac {\partial s}{\partial e}
+ \varepsilon_e^2 \nabla^2 e \right) q_n \nonumber\\
\nonumber\\
\nonumber\\
& &
+ \varepsilon_x^2 
\left(\frac{\partial X}{\partial t}\right) \nabla_n X + \left (\frac {\partial s}{\partial X}
+ \varepsilon_x^2 \nabla^2X \right) J_n \nonumber\\ 
\nonumber\\ 
\nonumber\\
& &
+ \varepsilon_\phi^2 \left(\frac{\partial \Phi}{\partial t}\right) \nabla_n \Phi   
\label{flux} 
\end{eqnarray} 
is the projection of the entropy flux vector on the normal vector $\vec n$, 
and 
\begin{eqnarray}
& & \sigma_S = 
\vec q \cdot \left [\nabla \left (\frac {\partial s}{\partial e} 
+ \varepsilon_e^2 \nabla^2 e \right) - 
\alpha_q \frac {\partial \vec q}{\partial t}\right] \nonumber\\
\nonumber\\ 
\nonumber\\
& &
+ \vec J \cdot \left [\nabla \left (\frac {\partial s}{\partial X}
+\varepsilon_x^2 \nabla^2 X \right)-
\alpha_j \frac {\partial \vec J}{\partial t}\right] \nonumber\\
\nonumber\\ 	\nonumber\\  
& & 
+ \frac{\partial \Phi}{\partial t} \left[ \frac{\partial s}{\partial \Phi} + 
\varepsilon_\phi^2 \nabla^2 \Phi  - 
\alpha_\phi \frac{\partial^2 \Phi}{\partial t^2} \right] > 0
\label{prod}
\end{eqnarray} 
is the local entropy production which has a bilinear form in the fluxes 
($\vec q$, $\vec J$, and $\partial \Phi/\partial t$) and their respective conjugate forces 
(expressions inside the square brackets). 

\subsection{Governing equations and thermodynamic consistency}\label{sec:gov_equats} 
Relation (\ref{flux}) is well known from the phase-field model based 
on CIT (see, e.g., Ref.~\cite{bis}), 
whereas the entropy production (\ref{prod}) includes the additional terms 
$-\alpha_q \partial \vec q/\partial t$, 
$-\alpha_j \partial \vec J/\partial t$, and $-\alpha_\phi \partial^2 \Phi / \partial t^2$ 
related to the nonequilibrium part of the 
generalized entropy. This is due to the special form for entropy, 
Eq.~(\ref{0}), and has a clear physical meaning: far from equilibrium, 
the dissipative fluxes provide ordering that leads to a decrease of 
the entropy production near a steady state 
as compared with the local-equilibrium state 
characterized by the same values of $e$, $X$, and $\Phi$. 

The production $\sigma_S$ of the 
generalized entropy, Eq.~(\ref{0}) is  
positive due to the statement of the second law of thermodynamics. 
This condition implies a relation between fluxes and forces which, 
in the simplest cases, is assumed to be linear. For Eq.~(\ref{prod}), 
it gives the following set of equations: \\
- \textit{evolution equations for heat and solute diffusion fluxes} 
\begin{equation}
\left\{ \begin{array}{ll}
\vec q \\\\
\vec J 
\end{array} \right\} 
= \left( 
\mathcal{M}
\right)
\left\{ \begin{array}{ll}
\nabla \left (\displaystyle 
\frac {\partial s}{\partial e} 
+ \varepsilon_e^2 \nabla^2 e \right) - 
\alpha_q \displaystyle 
\frac {\partial \vec q}{\partial t} \\\\
\nabla \left (\displaystyle 
\frac {\partial s}{\partial X} 
+ \varepsilon_x^2 \nabla^2 X \right) - 
\alpha_j \displaystyle 
\frac {\partial \vec J}{\partial t} 
\end{array} \right\}, 
\label{26cab}
\end{equation}
- \textit{evolution equation for the phase-field}
\begin{eqnarray}
\frac{\partial \Phi}{\partial t} = M_\phi \left( \frac{\partial s}{\partial \Phi} + 
\varepsilon_\phi^2 \nabla^2 \Phi -\alpha_\phi \frac{\partial^2 \Phi}{\partial t^2}
\right), 
\label{101}
\end{eqnarray} 
where 
\begin{equation}
(\mathcal{M}) 
= \left( \begin{array}{cc} 
M_{ee} & M_{ex} \\\\
M_{xe} & M_{xx} 
\end{array} \right)  
\label{26caba}
\end{equation}
is the matrix of mobilities for thermal and solutal transport, and 
$M_\phi$ is the mobility of the diffuse interface.   
Dependent on composition, the interface mobility is assumed to be 
\begin{eqnarray}
M_\phi =(1-X)M_\phi^A+XM_\phi^B > 0, 
\label{mobil_001}
\end{eqnarray} 
where $M_\phi^A$ and $M_\phi^B$ are the interface mobility for the transformation 
in pure systems consisting of $A$ or $B$ components, respectively. In various 
formulations of the phase-field model \cite{wheel201,karma1}, the mobilities of 
$M_\phi^A$ and $M_\phi^B$ are proportional to the atomic interface kinetic 
coefficient $\mu_0$ and inversely proportional to the interface width $W_0$, 
so that $M_\phi \sim \mu_0/W_0$. 

The matrix (\ref{26caba}) of transport and the interface mobility (\ref{mobil_001})
are assumed to be positively defined for the 
positive entropy production $\sigma_S$. 
The matrix (\ref{26caba}) can be considered as symmetric, so that the 
matrix can be regarded as being positive with the inequality: 
$M_{ee} M_{xx}>M_{ex}^2$. 
Note that the linear phenomenological laws given by Eqs.~(\ref{26cab}) and (\ref{101}) 
adopt the representation theorem of isotropic tensors \cite{true1}
according to which fluxes and forces of different tensorial rank do not couple 
as far as linear relations are involved (this independence of processes of different 
tensorial rank is also known as the Curie principle). In our case, the vectors of heat and 
solute diffusion fluxes cannot give rise to the flux of the scalar phase-field flux 
in a linear description. More complicated nonlinear relations 
between fluxes and forces consistent with positive entropy 
production in EIT are considered elsewhere \cite{j1,j,j2}. 

For simplicity, we ignore both kinds of 
``cross coupling'' effects in Eq.~(\ref{26cab}), so that 
$M_{ex}=M_{xe}=0$. Then, substitution of the fluxes from 
Eq.~(\ref{26cab}) into the balances (\ref{B511}), 
respectively, gives \\
- \textit{the governing equation for energy density}
\begin{eqnarray}
\tau_T \frac{\partial^{2} e}{\partial t^{2}} + 
\frac{\partial e}{\partial t} = 
- \nabla \cdot \Bigl[ M_{ee} \nabla \left (\frac {\partial s}{\partial e} 
+ \varepsilon_e^2 \nabla^2 e \right) \Bigr], 
\label{103abc}
\end{eqnarray} 
- \textit{the governing equation for solute concentration}
\begin{eqnarray}
\tau_D \frac{\partial^{2} X}{\partial t^{2}} + 
\frac{\partial X}{\partial t} = 
 - \nabla \cdot \Bigl[ M_{xx} \nabla \left (\frac {\partial s}{\partial X} 
+ \varepsilon_x^2 \nabla^2 X \right) \Bigr], 
\label{103bac}
\end{eqnarray} 
in which $\tau_T=\alpha_q M_{ee}$ is the relaxation time for the heat diffusion flux,  
and $\tau_D=\alpha_j M_{xx}$ is the relaxation time for solute diffusion 
(see Eqs.~(\ref{A1}) and Table \ref{tab:1}). 
After simplifying the transformation, Eq.~(\ref{101}) leads to \\ 
- \textit{the governing equation for the phase-field}
\begin{eqnarray}
\tau_{\Phi} \frac{\partial^2 \Phi}{\partial t^2} + \frac{\partial \Phi}{\partial t} 
= M_\phi \left( \frac{\partial s}{\partial \Phi} + 
\varepsilon_\phi^2 \nabla^2 \Phi \right),  
\label{101_a}
\end{eqnarray} 
where $\tau_{\Phi}=\alpha_\phi M_\phi$ is the timescale of the phase-field kinetics. 
According to Eq.~(\ref{101_a}), the acceleration $\partial^2 \Phi /\partial t^2$ of the phase-field 
appears due to introduction of both $\Phi$ and $\partial \Phi/ \partial t$ as 
independent variables and characterizes inertial effects inside the width 
of diffuse interface. 

Equations (\ref{103abc})-(\ref{101_a}) are the central outcome of our proposal 
[or, to mention a more complicated setting, we could also refer to equations 
(\ref{26cab})-(\ref{101})]. 
The role of the relaxation times is clear: they characterize the delay with which 
$\vec q$ and $\vec J$ 
reduce to their classical expressions (corresponding to classical transport equations), 
and the delay with which the inertial effects in the dynamics of the interfacial 
region are lost. 
The relaxation terms may be neglected in many circumstances, 
but become crucial in some important situations, leading, for instance, 
to a maximum possible value for the speed of advancement of the interface 
(in contrast to classic theory which allows for an infinite speed of propagation), 
and to the possibility of oscillatory phenomena in the width of the interface. 
Thus, the role of the new terms is not simply to add some new undetermined parameters 
(relaxation times) allowing for an improved fit of experimental results. These terms also play 
an important conceptual role, as they drastically change the possible kinds of 
behavior of the system. 

Some comments on the consistency of our proposal can be outlined. 
First of all, we may refer to its internal consistency as a thermodynamic 
(macroscopic) theory. Second, one must check its consistency with microscopic 
descriptions based, for instance, on kinetic theory, or on linear response theory, 
or in other statistical (microscopic) theories. Finally, one must check its consistency 
with experimental results.

Here, we comment on the internal thermodynamic consistency and, 
in the next Section, we shall refer to its consistency with a statistical theory, 
based on the fluctuation-dissipation theorem. In this theoretical paper we 
do not refer to experimental results. We assume that a consistent 
nonequilibrium thermodynamic theory should satisfy two main conditions: \\
($i$) the generalized or extended entropy must be maximum at equilibrium; \\
($ii$) the entropy production must be positive. 

To these two conditions one 
could add two more requirements: \\ ($iii$) the second differential of the entropy 
with respect to its basic variables (which is related to the dynamics of 
the variables) must be negative in order to lead to dynamically stable 
solutions; \\ ($iv$) the generalized equations of state obtained by differentiation 
of the generalized entropy must have a physical meaning consistent with experiments.

It can be seen immediately that the essential conditions ($i$) and ($ii$) are satisfied 
in our proposal. Indeed, the form (\ref{0}) and (\ref{1}) of the entropy 
guarantees that homogeneous equilibrium state has the maximum 
entropy as compared to nonequilibrium states with the same local 
values of $e$, $X$ and $\Phi$. Furthermore, introduction of the constitutive 
equations (\ref{26cab})-(\ref{101}) into the expression (\ref{prod}) 
of the entropy production 
yields for the latter a definite positive expression: 
\begin{eqnarray}
\sigma_S = 
\left(\vec q, \vec J \right)
\left( 
\mathcal{M}
\right)^{-1}
\left\{ \begin{array}{ll}
\vec q \\\\
\vec J 
\end{array} \right\} 
+ M_\phi^{-1} \left( \frac{\partial \Phi}{\partial t} \right)^2 
> 0.
\label{consist}
\end{eqnarray} 
As we noted, the transport matrix 
$(\mathcal{M})$ and the interface mobility $M_\phi$ 
are assumed to be positively defined for the 
positive entropy production, $\sigma_S > 0$. 
If one included 
higher-order nonlinear terms into the entropy (\ref{0}) or in the constitutive 
equations (\ref{26cab})-(\ref{101}), thermodynamic consistency would be more 
difficult to check than in our second-order approximation (\ref{0}). 
This approximation is sufficient to deal with a wide range of physical problems.

We shall not deal with conditions ($iii$) and ($iv$), which are subtler and 
typically involve nonlinear effects. For an indication of their 
analysis in some situations involving only $\vec q$ as nonequilibrium 
variables, the reader is referred to the monograph \cite{j}.

\section{Generalization of the model}\label{sec:generalizat}
The governing equations (\ref{103abc})-(\ref{101_a}) present causal 
propagation of heat and mass signals and a dissipative-wave advancing of 
a diffuse interface. 
We generalize them into evolution equations which are nonlinear in time. 
First, equations of state are considered 
from the point of view of the relaxation functions for the fluxes. Second, 
nonlinear evolution equations of a general type are derived from a variational 
formulation.

\subsection{Relaxation functions for the fluxes}\label{sec:relax_funct} 
Let's take into consideration a prehistory of the change 
of the phase-field in a point of a system. Such a prehistory must be 
taken if the system is not in local equilibrium. We shall use a functional 
description with a memory function.  

We use the entropy functional (\ref{1}), as before, 
to derive the equations of the model. 
In the absence of local equilibrium, one may incorporate the prehistory of the diffusion 
process. Then, the connections between the fluxes, $\vec q$, $\vec J$, 
and $\partial \Phi/\partial t$, from the one side, and driving forces, 
$\nabla (\delta S/\delta e)$, $\nabla (\delta S/\delta X)$, and $\delta S/\delta \Phi$, 
from the other side, are defined by the following integral forms: \\
- \textit{relaxation of the heat flux}  
\begin{eqnarray}
\vec q(\vec r,t) = \int_{-\infty}^{t} 
D_q(t-t^*)\nabla \frac{\delta S(t^*,\vec r)}{\delta e}dt^*,
\label{991A}
\end{eqnarray}
- \textit{relaxation of the solute diffusion flux}  
\begin{eqnarray}
\vec J(\vec r,t) = \int_{-\infty}^{t} 
D_j(t-t^*)\nabla \frac{\delta S(t^*,\vec r)}{\delta X}dt^*, 
\label{991}
\end{eqnarray}
- \textit{relaxation of the phase-field rate of change}  
\begin{eqnarray}
\frac{1}{M_\phi}\frac{\partial \Phi(\vec r,t)}{\partial t}
= - \int_{-\infty}^{t} 
D_\phi(t-t^*) \frac{\delta S(t^*,\vec r)}{\delta \Phi}dt^*,
\label{991B}
\end{eqnarray}
where  
$D_R=\{ D_q,D_j,D_\phi \}$ are the relaxational kernels for the fluxes, 
and the variational derivatives are obtained from the following expressions
\begin{eqnarray}
& & \frac{\delta S}{\delta e}=\frac {\partial s}{\partial e}
+ \varepsilon_e^2 \nabla^2e, \qquad  
\frac{\delta S}{\delta X}=\frac {\partial s}{\partial X}
+ \varepsilon_x^2 \nabla^2X, \nonumber\\
\nonumber\\
& & \frac{\delta S}{\delta \Phi}=\frac {\partial s}{\partial \Phi} 
+ \varepsilon_\phi^2 \nabla^2\Phi.   
\label{01} \end{eqnarray}

After substitution of expressions for the heat flux relaxation, 
Eq.~(\ref{991A}), and the solute diffusion relaxation, Eq.~(\ref{991}),
into the balance laws for energy and solute concentration, Eq.~(\ref{B511}), 
respectively, one can get the following integro-differential equations  
\begin{eqnarray}
& & \frac{\partial e(\vec r,t)}{\partial t} = 
 -\nabla \cdot \int_{-\infty}^{t}
D_q(t-t^*)\nabla \frac{\delta S(t^*,\vec r)}{\delta e}dt^*, \nonumber\\
\nonumber\\
& & \frac{\partial X(\vec r,t)}{\partial t} = 
 -\nabla \cdot 
\int_{-\infty}^{t}  
D_j(t-t^*)\nabla \frac{\delta S(t^*,\vec r)}{\delta X}dt^*. 
\label{103}
\end{eqnarray} 
Together with relaxation of the phase-field, Eq.~(\ref{991B}), 
the general system evolution 
during phase transformation is described by Eqs.~(\ref{103}). 

When the relaxation functions $D_R$ 
are specially defined, Eqs.~(\ref{991B}) and (\ref{103}) can be 
reduced to known models. Particularly,  
for the important class of dissipative and hyperbolic models, 
one can take the relaxation kernels in the following forms 
\begin{equation}
D_R=
\left\{ 
\begin{array}{ll}
D_R(0)\equiv const, & \textrm{wave propagation}, \\\\
D_R(0)\delta(t-t^*), & \textrm{dissipation}, \\\\
D_R(0)\exp \Big(- \displaystyle \frac{t-t^*}{\tau}\Big), & \textrm{wave and dissipation},
\end{array} 
\right. 
\label{26u}
\end{equation}
\\ 
where $D_R(0)=\{ D_q(0),D_j(0),D_\phi(0) \}$ are the relaxational kernels for the fluxes 
at present time $t=t^*$, 
and $\tau=\{ \tau_T, \tau_D, \tau_\Phi \}$ are the characteristic relaxation times for the fluxes. 

The different transformations within the diffuse-interface are 
described by different kernels in the integrals (\ref{991A})-(\ref{991B}). 
As it follows from Eq.~(\ref{26u}), the relaxation 
functions $D_R$ describe the memory of the system by 
assigning different weights to different moments in the past. 
Dissipation corresponds to a zero-memory transformation, 
i.e. the only relevant contributions are the "last" ones. In 
contrast to this situation, the infinite memory transformation 
with $D_R\equiv const$ leads to undamped wave propagation of 
the heat, solute, or the interface advancement. In between, the 
combination of the wave and dissipative regimes described 
by the exponentional law can be observed during rapid phase 
transformations. This is the case of hyperbolic phase-field model 
described in Sec. \ref{sec:hyper_pfm}. For the latter case, 
the relevance of all contributions to the fluxes decreases as the 
system moves to the past. 

In Sec. \ref{sec:hyper_pfm}, the model macroscopic consistency 
of the statements of EIT has been shown. Now, the consistency of our macroscopic 
approach with a microscopic description is verified in relation 
to the outcomes following from the fluctuation-dissipation theorem. 

The memory functions introduced in Eqs.~(\ref{991A})-(\ref{991B}) 
may be related to our analysis of the dynamics of the fluxes $\vec q$ and 
$\vec J$ and of $\partial \Phi/\partial t$ proposed by constitutive 
equations (\ref{26cab})-(\ref{101}). To do this, first, we may consider the 
fluctuation-dissipation theorem relating response memory functions to 
the time-correlation function of the corresponding fluxes 
(see, e.g., Ref. \cite{balescu}). 
This will allow us to show the consistency of 
our macroscopic formulation with the microscopic basis provided 
by the fluctuation-dissipation theorem.

The corresponding expressions are
\begin{eqnarray}
& & D_q(t-t^*)=\frac{1}{k_BT^2} \langle \widehat{\vec q}(t) \widehat{\vec q}(t^*) \rangle _{eq}, \nonumber\\
\nonumber\\
& & D_j(t-t^*)=\frac{1}{k_BT} \langle \widehat{\vec J}(t) \widehat{\vec J}(t^*) \rangle _{eq}, \nonumber\\
\nonumber\\
& & D_\phi(t-t^*)=\frac{1}{k_BT} \langle \widehat{ \partial _t\Phi (t)} 
\widehat{ \partial _t \Phi (t^*)} \rangle _{eq}.  
\label{X}
\end{eqnarray} 
Here $k_B$ is Boltzmann's constant, $\vec q$, $\vec J$, and $\partial _t \Phi$ 
stand for the microscopic operators 
for the heat flux, diffusion flux and the time derivative of $\Phi$, respectively, 
and $\langle ... \rangle _{eq}$ means an average over an equilibrium  
ensemble in statistical mechanics (as, for instance, the canonical one).

Relations (\ref{X}) play an important role in modern statistical mechanics, 
and may be formally derived from the Liouville equation in the framework 
of linear-response theory or from information theory \cite{balescu,zub}. 
However, from a practical point of view, 
the computation of the evolution of the microscopic 
operators for $\vec q$, $\vec J$ or $\partial _t \Phi$ 
on purely microscopic grounds is an 
overwhelming task exceeding actual capabilities. 
Such an evolution is either obtained by computer simulations, 
or tentatively given by a reasonable form inspired on phenomenological 
grounds. Thus, our evolution equations (\ref{26cab})-(\ref{101}) for 
$\vec q$, $\vec J$ and $\partial _t \Phi$ 
may be considered as a macroscopic modelling of the evolution 
of the fluxes, which according to Eq.~(\ref{X})  
is equivalent to proposing a form 
for the corresponding memory functions introduced in Eqs.~(\ref{991A})-(\ref{991B}). 
In general terms, it could be said that, according to Eq.~(\ref{X}), the study 
of the evolution of the fluxes around equilibrium is equivalent 
to the determination of the corresponding memory functions.

Constitutive equations (\ref{26cab})-(\ref{101}) imply that fluctuations of 
$\vec q$ and $\vec J$ near a homogeneous equilibrium state will decay exponentially as
$\vec q(t)=\vec q(0)\exp (-t/\tau_T)$ and $\vec J(t)=\vec J(0)\exp (-t/\tau_D)$. 
Introducing these expressions into Eq.~(\ref{X}) one obtains
\begin{eqnarray}
& & D_q(t-t^*)=\frac{1}{k_BT^2} \langle \widehat{\vec q}(0) \widehat{\vec q}(0) \rangle _{eq}
\exp \left( - \frac{t-t^*}{\tau_T} \right), \nonumber\\
\nonumber\\
& & D_j(t-t^*)=\frac{1}{k_BT} \langle \widehat{\vec J}(0) \widehat{\vec J}(0) \rangle _{eq}
\exp \left( - \frac{t-t^*}{\tau_D} \right),  
\label{XY}
\end{eqnarray}
which may be rewritten as
\begin{eqnarray}
& & D_q(t-t^*)=D_q(0)\exp \left( - \frac{t-t^*}{\tau_T} \right), \nonumber\\
\nonumber\\
& & D_j(t-t^*)=D_j(0)\exp \left( - \frac{t-t^*}{\tau_D} \right).  
\label{Y} 
\end{eqnarray} 
Indeed, when the microscopic expressions for $\vec q$ 
and $\vec J$ 
corresponding to an ideal gas are introduced into Eq.~(\ref{Y}) 
and the equilibrium average is performed 
(over a Maxwell-Boltzmann distribution function), 
the results for the $D_q(0)$ and $D_j(0)$ are equivalent 
to those obtained from the kinetic theory of gases 
in the time-relaxation approximation \cite{balescu}.

Note, finally, that the usual transport coefficients 
(thermal conductivity, diffusion coefficient) may be 
obtained (when the relaxation time is sufficiently short) 
by integration of Eq.~(\ref{X}), as
\begin{eqnarray}
& & \lambda = \frac{1}{k_BT^2} \int_{-\infty}^{\infty}  
\langle \widehat{\vec q}(t) \widehat{\vec q}(0) \rangle _{eq} dt, \nonumber\\
\nonumber\\
& & D = \frac{1}{k_BT} \int_{-\infty}^{\infty} 
\langle \widehat{\vec J}(t) \widehat{\vec J}(0) \rangle _{eq} dt,  
\label{Z}
\end{eqnarray}
which are the well-known Green-Kubo formulae for transport coefficients 
\cite{j,zub,balescu}. 
Thus, our macroscopic formalism is consistent with the microscopic 
fluctuation-dissipation theorem. It provides, in fact, a phenomenological 
complement to the fluctuation-dissipation expressions, which are the formal 
expressions from which it is difficult to obtain on exact grounds 
the form of the memory functions.

\subsection{A variational principle and Euler-Lagrange equations}\label{sec:tsv}
We assume, as above, that the generalized entropy density $s$ is a continuous 
and differentiable function defined by the local equilibrium contribution $s_e$ 
and flux-dependent nonequilibrium part $s_{ne}$ with the total set of variables 
(\ref{A3}) and generalized Gibbs equation (\ref{B21}). 
The balance equations for the heat and solute are the same, Eqs.~(\ref{B511}), 
and the local evolution of the entropy density is described by Eqs.~(\ref{15}).

A generalization can be given by introducing the generalized terms 
for derivatives into the entropy density with respect to classical 
variables ($e,X,\Phi$) and fluxes ($\vec q,\vec J,\partial \Phi/\partial t$), 
and also by introducing general forms of the entropy flux $\vec J_S$ and the source $\sigma_S$ 
in Eq.~ (\ref{15}). Depending on their own tensorial character, these are 
\begin{eqnarray} & & 
\left( \frac{\partial s}{\partial e} \right)_{\vec q} 
= \beta_1^e(e,I_q), \qquad \left( \frac{\partial s}{\partial X} \right)_{\vec J} 
= \beta_1^X(X,I_j), \nonumber\\ 
\nonumber\\
& & \left( \frac{\partial s}{\partial \Phi} \right)_\frac{\partial \Phi}{\partial t} 
= \beta_1^\Phi(\Phi,I_\frac{\partial \Phi}{\partial t}), \qquad
\left( \frac{\partial s}{\partial \vec q} \right)_e = \beta_2^e(e,I_q) \vec q, 
\nonumber\\
\nonumber\\ 
& & \left( \frac{\partial s}{\partial \vec J} \right)_X = \beta_2^X(X,I_j) \vec J, \nonumber\\
\nonumber\\ 
& & \left( \frac{\partial s}{\partial (\partial \Phi/\partial t)} \right)_\Phi 
= \beta_2^\Phi (\Phi,I_\frac{\partial \Phi}{\partial t}) \frac {\partial \Phi}{\partial t}, \nonumber\\ 
\nonumber\\ 
& & \vec J_S = \beta_3^e(e,I_q) \vec q 
+ \beta_3^X(X,I_j) \vec J 
+ \beta_3^\Phi(\Phi,I_\frac{\partial \Phi}{\partial t}) \frac{\partial \Phi}{\partial t}, \nonumber\\
\nonumber\\
& & \sigma_S= \beta_4^e(e,I_q) + \beta_4^X(X,I_j) + \beta_4^\Phi(\Phi,I_\frac{\partial \Phi}{\partial t}),
\label{17u} 
\end{eqnarray}
where 
\begin{eqnarray}
I_q=\vec q \cdot \vec q, \qquad 
I_j=\vec J \cdot \vec J,  \qquad
I_\frac{\partial \Phi}{\partial t}=\left( \frac{\partial \Phi}{\partial t}\right)^2
\label{20} 
\end{eqnarray} 
are the single scalar invariants of the extended set 
(\ref{A3}) of variables, and $\beta_i$ are the scalar functions depending on classic variables ($e,X,\Phi$) 
and invariants $I_i$. Then, utilizing Eqs.~(\ref{17u}), the generalized Gibbs equation (\ref{B21}) 
gives the time derivative of the entropy density as follows: 
\begin{eqnarray}
& & \frac{\partial s}{\partial t}  
= \beta_1^e(e,I_q)\frac{\partial e}{\partial t} + \beta_2^e(e,I_q) 
		\vec q \cdot 
			\frac{\partial \vec q}{\partial t} \nonumber\\ 
\nonumber\\ 
& & + \beta_1^X(X,I_j)\frac{\partial X}{\partial t} 
+ \beta_2^X(X,I_j) \vec J \cdot \frac{\partial \vec J}{\partial t}  \nonumber\\ 
\nonumber\\ 
& &  + \beta_1^\Phi(\Phi,I_\frac{\partial \Phi}{\partial t}) \frac{\partial \Phi}{\partial t}
+ \beta_2^\Phi (\Phi,I_\frac{\partial \Phi}{\partial t}) 
\frac {\partial \Phi}{\partial t} \frac{\partial^2 \Phi}{\partial t^2}. 
\label{18c} 
\end{eqnarray}

Locally, Eq.~(\ref{15}) is satisfied as a balance law and, for the entire system, one can 
postulate extremal condition in the Lagrangian form of 
$\mathcal{L}=\int_v (\partial s/\partial t + \nabla \cdot \vec J_S - \sigma_S) dv \rightarrow extr$, 
implying an extremal difference between the ``kinetic'' part 
$\int_v (\partial s/\partial t + \nabla \cdot \vec J_S) dv $  
and the ``potential'' part $\int_v \sigma_S dv$ 
for the whole nonequilibrium system. Then, the entropy density 
satisfies the following variational principle \cite{mexican1}
\begin{eqnarray}
\delta \mathcal{L}= 
\delta \int_v dv \left(\frac{\partial s}{\partial t} + \nabla \cdot \vec J_S - \sigma_S \right) =0, 
\label{18a} 
\end{eqnarray}
in which the variation $\delta$ is carried out only on the nonconserved flux variables 
$\vec q$, $\vec J$, and $\partial \Phi / \partial t$, 
i.e. $\delta$ is taken only over the space $\{F\}$ 
from the set (\ref{A3}) while the variables $e$, $X$, and $\Phi$ 
from the set $\{C\}$ remain constant during the variation. 
Also, during the variation, the tangent thermodynamic space 
[time and spatial derivatives from the set (\ref{A3})] is fixed. From this it 
follows that Eq.~(\ref{18a}) is a variational principle of a restricted type.

Using balance laws (\ref{B511}), 
substitution of Eqs.~(\ref{17u}) and (\ref{18c}) into 
variational principle (\ref{18a}) leads to    
\begin{eqnarray}
& & \delta \int_v dv \bigg[ (\beta_3^e-\beta_1^e)
							\nabla \cdot \vec q + 
\left( \beta_2^e\frac{\partial \vec q}{\partial t} 
+ \nabla \beta_3^e \right) \cdot \vec q - \beta_4^e \nonumber\\ 
& & + (\beta_3^X-\beta_1^X) \nabla \cdot \vec J + 
\left( \beta_2^X\frac{\partial \vec J}{\partial t} + \nabla \beta_3^X \right) 
\cdot \vec J - \beta_4^X  \nonumber\\ 
\nonumber\\ 
& & 
+ \beta_1^\Phi \frac{\partial \Phi}{\partial t} 
+ \beta_3^\Phi \nabla \frac{\partial \Phi}{\partial t} + 
\left( \beta_2^\Phi\frac{\partial^2 \Phi}{\partial t^2} + \nabla \beta_3^\Phi \right) 
\frac{\partial \Phi}{\partial t} - \beta_4^\Phi \bigg] = 0.  \nonumber\\ 
\label{18b}
\end{eqnarray}
Variation of Eq.~(\ref{18b}) is obtained by taking as constants the time 
derivatives, gradients and divergences. Since $\delta I_q = 2\vec q \cdot \delta \vec q$,  
$\delta I_j = 2\vec J \cdot \delta \vec J$, and 
$\delta I_\frac{\partial \Phi}{\partial t} 
= 2(\partial \Phi/\partial t)\delta (\partial \Phi/\partial t)$
from Eq.~(\ref{18b}) one gets 
\begin{widetext}
\begin{eqnarray}
& & \int_v dv \bigg[ 2\left( \frac{\partial \beta_3^e}{\partial I_q} 
		- \frac{\partial \beta_1^e}{\partial I_q} \right) 
			\vec q(\nabla \cdot \vec q) 
				+ \beta_2^e \frac{\partial \vec q}{\partial t} 
+ 2\frac{\partial \beta_2^e}{\partial I_q}\vec q\vec q \cdot 
					\frac{\partial \vec q}{\partial t} 
+ \nabla \beta_3^e 
		- 2\frac{\partial \beta_4^e}{\partial I_q}\vec q
				\bigg] \delta \vec q \nonumber\\ 
\nonumber\\ & & 
+ \int_v dv \bigg[ 2\left( \frac{\partial \beta_3^X}{\partial I_j} 
	- \frac{\partial \beta_1^X}{\partial I_j} \right) 
		\vec J(\nabla \cdot \vec J) 
+ \beta_2^X \frac{\partial \vec J}{\partial t}
+2\frac{\partial \beta_2^X}{\partial I_j}\vec J\vec J \cdot \frac{\partial \vec J}{\partial t} 
+ \nabla \beta_3^X - 2\frac{\partial \beta_4^X}{\partial I_j}\vec J
\bigg] \delta \vec J  \nonumber\\ 
\nonumber\\
& &  + \int_v dv \bigg[ 
2\frac{\partial \beta_1^\Phi}{\partial I_\frac{\partial \Phi}{\partial t}}  
\left(\frac{\partial \Phi}{\partial t}\right)^2 
+ 2\frac{\partial \beta_3^\Phi}{\partial I_\frac{\partial \Phi}{\partial t}} 
\frac{\partial \Phi}{\partial t}\nabla\frac{\partial \Phi}{\partial t}
+ \beta_2^\Phi \frac{\partial^2 \Phi}{\partial t^2} 
+ 2\frac{\partial \beta_2^\Phi}{\partial I_\frac{\partial \Phi}{\partial t}}\frac{\partial^2 \Phi}{\partial t^2}
\left(\frac{\partial \Phi}{\partial t}\right)^2 + \nabla \beta_3^\Phi 
- 2\frac{\partial \beta_4^\Phi}{\partial I_\frac{\partial \Phi}{\partial t}} 
\frac{\partial \Phi}{\partial t} \bigg] 
\delta \left(\frac{\partial \Phi}{\partial t} \right)= 0. \nonumber\\ 
\nonumber\\ 
\label{19}
\end{eqnarray}
\end{widetext}
Due to arbitrary variation of $\delta \vec q$, 
$\delta \vec J$, and $\delta (\partial \Phi/\partial t)$, 
the Euler-Lagrange equations 
directly follow from Eq.~(\ref{19}). These are \\
- \textit{evolution equation for the heat flux}  
\begin{eqnarray} 
& & \left( \frac{\partial \beta_2^e}{\partial I_q}\vec q\vec q + \beta_2^e \mathcal{U} 
\right)\cdot \frac{\partial \vec q}{\partial t}  \nonumber\\
\nonumber\\
& & + \left[ \left( \frac{\partial \beta_3^e}{\partial I_q} - \frac{\partial \beta_1^e}{\partial I_q} \right) 
\nabla \cdot \vec q - \frac{\partial \beta_4^e}{\partial I_q} \right] \vec q =-\frac{1}{2} \nabla \beta_3^e, 
\label{200}
\end{eqnarray} 
- \textit{evolution equation for the solute diffusion flux} 
\begin{eqnarray}
& & \left( \frac{\partial \beta_2^X}{\partial I_j}\vec J\vec J 
+ \beta_2^X \mathcal{U} 
\right)\cdot \frac{\partial \vec J}{\partial t} \nonumber\\ 
\nonumber\\
& & + \left[ \left( \frac{\partial \beta_3^X}{\partial I_j} - \frac{\partial \beta_1^X}{\partial I_j} \right)
\nabla \cdot \vec J - \frac{\partial \beta_4^X}{\partial I_j} \right] \vec J =-\frac{1}{2} \nabla \beta_3^X, \nonumber\\ 
\label{21} 
\end{eqnarray} 
- \textit{evolution equation for the phase-field} 
\begin{eqnarray} & & 
\left( \frac{\partial \beta_2^\Phi}{\partial I_\frac{\partial \Phi}{\partial t}}
\left(\frac{\partial \Phi}{\partial t}\right)^2 + \frac{1}{2}\beta_2^\Phi \right)
\frac{\partial^2 \Phi}{\partial t^2}  \nonumber\\
\nonumber\\
& & + \left[  \frac{\partial \beta_3^\Phi}{\partial I_\frac{\partial \Phi}{\partial t}}\nabla \frac{\partial \Phi}{\partial t}
+ \frac{\partial \beta_1^\Phi}{\partial I_\frac{\partial \Phi}{\partial t}}\frac{\partial \Phi}{\partial t} 
- \frac{\partial \beta_4^\Phi}{\partial I_\frac{\partial \Phi}{\partial t}} \right] \frac{\partial \Phi}{\partial t} 
= - \frac{1}{2} \nabla \beta_3^\Phi, \nonumber\\
\label{22}
\end{eqnarray} 
where $\mathcal{U}$ is the unit tensor of second rank.

Eqs.~(\ref{200})-(\ref{22}) are nonlinear evolution equations for  
$\vec q$, $\vec J$, and $\partial \Phi/\partial t$ and they are of the general 
form of evolution equations (\ref{26cab})-(\ref{101}). Indeed, the nonlinearity 
is clearly seen from the following form of these equations: 
\begin{eqnarray} 
& & 
\tau_T(e,\vec q) \frac{\partial \vec q}{\partial t}+\vec q=M_{ee}(e,\vec q)\nabla \beta_3^e, \nonumber\\
\nonumber\\
& & \tau_D(X,\vec J) \frac{\partial \vec J}{\partial t}+\vec J=M_{xx}(X,\vec J)\nabla \beta_3^X, 
\nonumber\\ 
\nonumber\\ 
& & \tau_\Phi 
\left(\Phi, \frac{\partial \Phi}{\partial t}\right) 
\frac{\partial^2 \Phi}{\partial t^2} + 
\frac{\partial \Phi}{\partial t} = M_\Phi \left(\Phi, \frac{\partial \Phi}{\partial t} \right)\nabla \beta_3^\Phi, 
\label{22_a}
\end{eqnarray} 
where $\tau_i$ and $M_i$ are the functions of the classic set $\{C\}=\{e,X,\Phi\}$ 
as well as nonlinear functions of the fluxes 
which can be explicitly found from Eqs.~(\ref{200})-(\ref{22}) and relations 
(\ref{17u})-(\ref{20}). Thus, taking the generalized evolution of the entropy density, 
Eq.~(\ref{18c}), and using variational principle (\ref{18a}), 
we arrive to nonlinear general evolution equations for fluxes, Eqs.~(\ref{22_a}), 
which might be merely reduced to the evolution equations 
(\ref{26cab})-(\ref{101}) of the hyperbolic phase-field model. 

\section{Relation to existing models}\label{sec:comparison}
It is interesting to note that sharp-interface and diffuse-interface 
models with relaxation of fluxes have been used to describe transient processes 
in various nonequilibrium systems (see monograph 
\cite{temam1}, Chapter 4). 
Therefore we synthesize here several previous and 
very recent results in comparison with the 
developed hyperbolic model (Sec. \ref{sec:hyper_pfm}) 
and generalized model (Sec. \ref{sec:generalizat}) of rapid 
phase transformation. 

\subsection{Superconductivity}
Ginzburg and Landau established their variational principle for 
the continuous transition from the normal to the 
superconducting phase \cite{ginz}. They used a free energy density with 
a gradient term which has been further used in many phenomena 
(e.g., in description of spinodal decomposition 
\cite{ch1} or crystal growth \cite{ch2}). As a logical extension, 
the transition between the normal and the superconducting phases 
can be described with the delay given by equations of the hyperbolic model 
[starting from the functional of the form (\ref{1})] or using generalized 
models with memory, Sec. \ref{sec:relax_funct}. 

Generally, equations (\ref{103abc})-(\ref{101_a}) are consistent with 
the generalized entropy density given by Eq.~(\ref{0}). The equations are reduced 
to the classic equations from Refs. \cite{ginz,ch1,ch2} when the times 
$\tau_T$, $\tau_D$, and $\tau_\Phi$ tend to zero. 
Furthermore, the entropy density (\ref{0}) together with the evolution 
equations (\ref{26cab}) has been justified microscopically 
\cite{j,mr} for the one-component system and from Grad's 
procedure for monatomic gases. 

The choice of thermodynamic potential is important, as it governs the transition 
from metastable state to the stable one. 
Normally, the potential for transition is included in the 
expression for entropy density (or for free energy density) 
in the form of a double-well function or by a monotonically 
increasing function incorporating nonequilibrium conditions 
at the interface \cite{ginz,sahm,bragard}. 
In the present paper, 
we do not give an explicit form of $s_e$ in Eq.~(\ref{0}) 
and present governing equations (\ref{103abc})-(\ref{101_a}) 
[or variational derivatives (\ref{01})] in a general form. 
The choice of the thermodynamic potential might be given 
for the problem under consideration. 

\subsection{Glass transition, structural relaxation and phase separation}
J\"ackle et al. \cite{binder1} considered isothermal phase transformation 
in the presence of additional slow structural relaxation variables. 
Considering the dynamics based on the relaxational chemical potential, 
these authors refer their model to systems with phase 
separation and to slow structural relaxation in polymeric solutions 
in the proximity of the glass transition temperature. The calculation 
has shown that, even at the early stages of phase separation, equation 
for chemical potential with memory may give  
pronounced deviations from the predictions of classic 
linear Cahn-Hillard's model \cite{ch1}. 

Phase separation during spinodal decomposition may proceed  
under local nonequilibrium conditions in solute diffusion field 
offered by rapid quenching. As it has been demonstrated in computational 
modeling \cite{bastea1}, the rapidly quenched liquid mixtures under 
decomposition exhibit non-equilibrium patterns, evolving 
with universalities different from those extracted 
from the Cahn and Hillard's model. 

Local nonequilibrium separation in liquids can be described in terms of EIT 
as a model for isothermal spinodal decomposition in a binary system \cite{g1}, 
in conditions of large deviation from thermodynamic equilibrium. 
The dynamics of the diffusion flux $\vec J$ as a fast variable from the set (\ref{A3}) 
is consistent with the characteristic time of local rearrangement of particles 
(atoms or molecules) or with the time of relaxation of diffusive flux 
to its local equilibrium steady-state value. 
The model equation for spinodal decomposition of a binary system is the 
generalized Cahn-Hillard equation of the form of Eq.~(\ref{103bac}) for 
local nonequilibrium solute redistribution. In this case, the dynamics 
of rapidly quenched decomposition is described for short periods of time 
or large gradients of chemical composition.  

\subsection{Shear flow, viscoelastic fluids and diffusion-reaction systems} 
The system of coupled evolution equations (\ref{26cab})-(\ref{101}) describes, 
in fact, a process of phase separation under shear if temperature 
is replaced by viscous pressure tensor. In this case, one may 
get the condition defining the spinodal line in non-equilibrium states 
[see monograph \cite{jou2000}, Chapter 6]. As a reduced one, 
equation of type (\ref{26cab}) or (\ref{103}), endowed with 
homogeneous Dirichlet boundary conditions, has been 
introduced to model the behavior of certain viscoelastic fluids 
and to predict the velocity of flow \cite{davis1}. 

In addition, equation of type (\ref{103}) 
is used to predict a wavefront in time-delayed reaction-diffusion systems 
of the generalized Fisher's equation \cite{fort1}. 
The speed of the travelling wave depends 
on the relaxation time and therefore spreading of population in reaction-diffusion system 
might be predicted with great flexibility. 
One of the consequences of this equation, 
reduced to Eqs. (\ref{103abc}) or (\ref{103bac}) for modelling a 
hyperbolic reaction-diffusion system  [with $\varepsilon_e=0$ or 
$\varepsilon_x=0$, respectively], might be considered in an 
exciting example suggested by Fort and Mendez in Ref.~\cite{fort2} 
for neolitic advancing of human groups across Europe. 
They have shown, in particular, that hyperbolic reaction-diffusion 
equations of type of Eqs.~(\ref{103abc}) or (\ref{103bac}) predict the population 
spreading during the European past in agreement with the existing 
archaeological data. 

\subsection{Rapid solidification}
At deep supercoolings in a 
solidifying system, or at high velocity of the solid-liquid interface, it is necessary 
to take into account local nonequilibrium effects in solute diffusion and to use a 
non-Fickian model for transport processes which is compatible with EIT 
\cite{gs,gd1}.  
The problem of rapid solidification within the sharp-interface limit is 
described by generalized Stefan problem (so-called ``self-consistent 
hyperbolic Stefan problem'' \cite{gd3,gd4}) 
which takes into account local 
nonequilibrium both at the interface and within the bulk phases. 
In such a case, the spatio-temporal evolution of solute 
concentration is described by the partial differential equation (\ref{103bac}) 
of a hyperbolic type [with $\varepsilon_x=0$] 
which takes into account the relaxation of solute diffusion 
flux to the local thermodynamical equilibrium in a 
rapidly solidifying system. 

Advancing of the diffuse-interface with a higher velocity comparable with 
the solute diffusion speed is also described by the phase-field 
model with relaxation of the diffusion flux \cite{g1}. It has been shown that 
choosing the concrete form of the entropy (as the thermodynamic potential), 
one may recover the existing models based on the CIT and analyze 
solidification under local nonequilibrium conditions. 

\subsection{Motion of antiphase domains}
In the description of diffuse interface kinetics, Allen and Cahn \cite{ch3} 
proposed a model for evolution of the non-conserved order field 
during antiphase domain coarsening. For isotropic interfaces, gradient 
flow gives the Allen-Cahn equation by taking $\tau_\Phi=0$ in Eq.~(\ref{101_a}). 
This equation is true in the case of low inertial effects 
in comparison with the dissipative effects. 
With finite relaxation time, $\tau_\Phi$, and 
finite acceleration, $\partial^2 \Phi/\partial t^2$, Eq.~(\ref{101_a}) 
predicts the evolution of coarsening with relaxation. 
It is reasonable to say that the generalized Allen-Cahn 
equation (\ref{101_a}) is true for the case of significant 
inertial effects during the motion of antiphase domains. 

As an advancing of the Allen and Cahn's model, 
the process of the interface motion by mean curvature 
with delayed response has been analyzed recently. 
Rotstein et al. \cite{r1} developed the phase-field model 
based on equations similar to Eqs.~(\ref{991A}) and (\ref{991B}). 
These authors described the first-order transition with the delayed response 
of the system due to slow relaxation of internal variables. Using the exponential 
relaxation function for wave and dissipative mode, Eq.~(\ref{26u}), 
which leads to the hyperbolic phase-field model, the dynamics of 
the perturbed motion of interface by mean curvature has been considered.  
It has been shown in Ref.~\cite{r1} that internal relaxing effects induce damped oscillations in 
the interfacial motion during crystalline coarsening. As opposed to the classic 
parabolic phase-field model, the hyperbolic phase-field model predicts these 
interfacial oscillations in qualitative consistency with the 
oscillations on the surface of quantum crystals \cite{andr1} and in crystallization 
waves in helium \cite{keshishev}. From a mathematical viewpoint, a search for  
existence and uniqueness of the solution and some 
well-posedness results for the problem of motion by mean curvature using the 
phase-field model with memory are beginning to be presented (see Ref.~\cite{r2}). 

\subsection{Complex (dusty) plasmas}
Recent investigation into the field of complex (dusty) plasma physics show that 
this system exhibits a complicated behavior which depends 
on the behavior of its ``subsystems'' which are represented by electrons, ions, 
neutral gas, and charged dust particles. 
All of them have their own relaxation time to local equilibrium; therefore, 
interaction among them may lead to a delay of relaxation 
to local equilibrium in plasma. 
Moreover, in the electronic subsystem of plasma, local 
equilibrium does not exist, that stimulates development 
of theories beyond local equilibrium \cite{tokatly}. 
Interaction of different subsystems in complex (dusty) 
plasmas with missing local thermodynamic equilibrium in 
the electronic subsystem, makes the description of observed experimental data 
of this object rather complicated.  

Experimental results of Morfill et al. \cite{morfill1} from 
plasma observations exhibit unusual behavior from 
weak collisionless interaction of gases to fluid flow 
with further possible crystallization of plasma. These results are 
described by means of molecular dynamic simulations 
\cite{morfill1}. The field approach, also, seems to be also 
applicable due to the fact that during transitions in plasma, 
the characteristic size of patterns is on the mesoscopic or even 
macroscopic scale. The field approach to a heat- and 
electronically-conducting fluid has been demonstrated in 
ionized gases \cite{mexican1} using  
equations of generalized type of Eqs.~(\ref{200})-(\ref{22}).

\section{Conclusions}\label{sec:con}
The diffuse-interface model for rapid phase transformation 
in metastable binary systems has been presented. 
To describe the steep but smooth change of phases  
within the width of diffuse interface, 
we use the formalism of the phase-field model. 

Rapid phase transformations may proceed under local 
nonequilibrium conditions. 
In our phenomenological macroscopic description, we 
extend the classic set of independent 
thermodynamic variables by inclusion of dissipative fluxes as 
additional basic variables. The evolution of the fluxes is 
characterized by their own dynamics with relaxation times 
$\tau$ summarized in Table \ref{tab:1}. 
Thus, the extended set (\ref{A3}) of variables allows one to 
describe phase transformations with finite interface velocity  
that is comparable or even higher than $l/\tau$, 
where $l$ is the mean-free-path of particles (atoms). 

The evolution equations for the hyperbolic phase-field model with 
dissipation are derived from an entropy functional (\ref{1}) based 
on the extended set (\ref{A3}) of independent thermodynamic 
variables. This model yields a definite positive entropy production 
(\ref{consist}) in consistency with the second law of thermodynamics. 

Generalization of the model has been done by introducing 
memory functions and using a variational principle. 
As a result, the consistency of the macroscopic approach 
with the microscopic fluctuation-dissipation theorem 
has been found for the phase-field with memory 
[Eqs. (\ref{X})-(\ref{Z})]; 
and nonlinear evolution equations [Eqs. (\ref{200})-(\ref{22})] 
are derived from the variational principle (\ref{18a}).  

The derived equations for the evolution of diffuse interface were 
correlated with existing models of nonequilibrium transport 
processes and for systems under phase transformation. 
Particularly, we compare our derivation to models of 
superconductivity, phase separation, viscoelastic or 
electronically-conducting fluids, 
interface motion by mean curvature, rapidly solidifying systems, 
and reaction-diffusion systems.

\begin{acknowledgments}
P.G. acknowledges financial support from the German Research Foundation 
(DFG - Deutsche Forschungsgemeinschaft) under the project No. HE 1601/13. 
He also acknowledges the support of the Administration of the Physical Statistics Group 
during his stay in Universitat Aut\`onoma de Barcelona. 
D.J. acknowledges financial support from the Direcci\`on General 
de Investigaci\`on of the Spanish Ministry of Science and 
Technology BFM 2003-06033 and the Direcci\`o General de 
Recerca of the Generalitat of Catalonia under grant 2001 SGR-00186. 
\end{acknowledgments}

\bibliography{basename of .bib file}

\begin{thebibliography}{99}
\bibitem{eval1}
Evaluated from Eq.~(\ref{A1}) for $\tau_T$ 
in which the thermal diffusivity 
$a=1.2 \cdot 10^{-5}$ ($m^2/s$) 
and the thermal speed 
$V_T=10^3$ ($m/s$) 
are accepted from Ref.~\cite{gd3}.
\bibitem{eval21}
Evaluated from Eq.~(\ref{A1}) for $\tau_D$ 
in which the diffusion constant $D=5.5 \cdot 10^{-9}$ ($m^2/s$) 
and the solute diffusion speed 
$V_D=18.9$ ($m/s$) are accepted from 
Ref.~\cite{gd1}.
\bibitem{eval211}
Evaluated from Eq.~(\ref{A1}) for $\tau_D$ 
in which the diffusion constant $D=3 \cdot 10^{-9}$ ($m^2/s$) 
and the solute diffusion speed 
$V_D=20$ ($m/s$) are accepted from 
Ref.~\cite{gd1}.
\end{thebibliography}

\end{document}